\documentclass[amsfonts,12pt]{article}

\advance\hoffset by -1.1truecm
\advance\voffset by -1.0truecm

\def\be{\begin{equation}}
\def\ee{\end{equation}}
\def\ba{\begin{eqnarray}}
\def\ea{\end{eqnarray}}

\newcommand{\N}{\mbox{\rm N} \hspace{-0.9em} \mbox{\rm I}\,\,\,}
\newcommand{\R}{\mbox{I \hspace{-0.82em} R}}

\newcommand{\x}{{\bf x}}

\newcommand{\sn}{\smallskip\newline $~~~~~$}
\newcommand{\mn}{\medskip\newline   $~~~~~$}

\def\one{1\hskip-.37em 1}
\def\E{{\rm I}\hskip-.2em{\rm E}}

\def\ra{\rightarrow}

\def\o{\overline}
\def\b{\begin{eqnarray*}}      
\def\e{\end{eqnarray*}}        
\def\bn{\begin{eqnarray}}         
\def\en{\end{eqnarray}}           
\def\<{\langle}
\def\>{\rangle}

\def\{{\lbrace}
\def\}{\rbrace}
\begin{document}
\title{On the Implementation of Constraints through Projection Operators}
\author{Achim Kempf and John R. Klauder
\\
Institute for Fundamental Theory\\
 Departments of Physics
and Mathematics\\
University of Florida, Gainesville, FL 32611, USA\\
{\small Email:  kempf@phys.ufl.edu, klauder@phys.ufl.edu}} 

\date{}

\maketitle

\vskip-7.5truecm

\hskip11.7truecm
{\tt UFIFT-HEP-00-24} 

\hskip11.7truecm
{\tt quant-ph/0009072}
\vskip7.1truecm

\begin{abstract}
Quantum constraints of the type
$Q\vert\psi_{phys}\rangle=0$ can be straightforwardly implemented 
in cases where $Q$ is a self-adjoint operator for which
zero is an eigenvalue. In that case, the physical Hilbert space
is obtained by projecting onto the kernel of $Q$,
i.e. $H_{phys}=\mbox{ker}~Q=\mbox{ker}~Q^*$.
It is, however, nontrivial to 
identify and project onto $H_{phys}$
when zero is not in the 
point spectrum but instead is
in the continuous spectrum of $Q$, because
then $\mbox{ker} ~Q=\emptyset$. 

Here, we observe that the
topology of the underlying Hilbert space can be harmlessly modified
in the direction perpendicular to the constraint
surface in such a way that $Q$ becomes non-self-adjoint.
This procedure then allows us to conveniently
obtain $H_{phys}$ as the proper Hilbert subspace 
$H_{phys}=\mbox{ker}~Q^*$
on which one can project as usual.
In the simplest case, the necessary change of topology amounts
to passing from an $L^2$ Hilbert space to a Sobolev space. 
\end{abstract}
\newpage
\section{Introduction}

Numerous classical dynamical systems are distinguished by the presence of 
constraints, which, in a phase space formulation, act to restrict the 
system to the {\it constraint hypersurface}, a submanifold of the 
original phase space with a positive co-dimension. In practice, 
this restriction is accomplished by one or more real constraint 
functions, $\phi_\alpha(p,q)=0$, $\alpha=1,2,\ldots,A$, $A<\infty$, 
which are nondynamical---no time derivatives---and serve to constrain
 the system to the constraint hypersurface. When dealing with concrete
 examples it is useful to further categorize a system of constraints
 (along with their associated Hamiltonian) into classes (first and 
second), as well as other subdivisions (closed, open, irreducible,
 reducible, regular, irregular, etc.); these categories are well 
described in the literature and are not reviewed here \cite{k1}.
 For purposes of the present paper, it is sufficient to focus on 
the several constraint functions themselves, and we need not be 
too concerned about any specific subclassification of the set of
 constraints. 

Our principal interest lies in quantization, or more particularly, 
with the quantum theory of constraints. We assume that as quantum 
operators, the constraints $\Phi_\alpha$, $\alpha=1,\ldots,A$, are
 represented by self-adjoint operators determined in some fashion 
{} from the classical constraint functions by some consistent but 
unspecified quantization procedure. Just as the classical constraints 
act to restrict the system to a subset of the original classical 
phase space, it is the role of the quantum constraints, in like 
manner, to restrict the system to a subset of the original quantum 
mechanical phase space. Since the ``quantum mechanical phase space''
 is a Hilbert space, such a restriction is ideally imposed by the
 several constraint conditions $\Phi_\alpha\,|\psi\>_{phys}=0$, 
$\alpha=1,\ldots,A$. As a linear equation, it follows that the 
vectors $|\psi\>_{phys}$ form a linear space, and since the 
constraint operators are self adjoint, the given space is closed,
 hence a subspace ${H}_{phys}$, the presumed {\it physical
 Hilbert space}, in the original Hilbert space, $H$. For 
evident reasons we focus attention on those cases where 
${\dim({H}}_{phys})>0$. 

The foregoing scenario does indeed hold for certain families
 of constraints, namely, in cases where the constraint operators 
$\Phi_\alpha$ each have a set of simultaneous eigenvectors with 
eigenvalue zero. When that is the case, we may also consider the
 {\it single} constraint operator 
$X^2\equiv\Sigma_{\alpha=1}^A\Phi_\alpha^2$, regarded as a
 self-adjoint operator, and observe that $X^2\,|\psi\>_{phys}=0$ 
is completely equivalent to the several equations 
$\Phi_\alpha\,|\psi\>_{phys}=0$, $\alpha=1,\ldots,A$. An example
 of this kind of constraint situation is given by 
$\Phi_\alpha=J_\alpha$, $\alpha=1,2,3$, where  the operators
 $J_\alpha$ satisfy the Lie algebra for SO(3) [or SU(2)] and 
the condition $\Sigma J^2_\alpha\,|\psi\>_{phys}=0$ corresponds 
 to a restriction to spherically symmetric states. We also observe 
that we can also set ${H}_{phys}\equiv\E{H}$, where 
$\E=\E^*=\E^2$ is a {\it projection operator}, which in the 
present case is $\E=\E(\Sigma J^2_\alpha=0)$. For comparison 
purposes with what is to come we also note, equivalently, that 
  \be \E=\E(\Sigma J^2_\alpha\le
 c\hbar^2)=\E(\Sigma J^2_\alpha=0)\;, \ee
where $c<2$ for SO(3) [or $c<3/4$ for SU(2)]. The discussion and
 example of the present paragraph refer to the ideal situation 
regarding the constraint operators.

More generally, the constraint operators do {\it not} fulfill the
 ideal (Dirac) criteria given above. In fact, it frequently 
happens that the set of constraint operators have no nonzero
 eigenvector with eigenvalue zero. This situation may arise 
in two fundamentally different ways. One of these ways refers
 to cases where $X^2\equiv\Sigma_{\alpha=1}^A\Phi^2_\alpha$ 
has a discrete spectrum (in the vicinity of zero) which {\it
 does not include zero}. An example of this situation is given
 by the two classical constraints $\phi_1=p=0$ and $\phi_2=q=0$
 (for the same degree of freedom), which then become the quantum 
constraints
 $\Phi_1=P$, and $\Phi_2=Q$, two operators which satisfy the
 Heisenberg commutation relation $[Q,P]=i$ (with $\hbar=1$). 
The ideal equations, $P\,|\psi\>_{phys}=0$ and $Q\,|\psi\>_{phys}=0$, 
imply (modulo domain issues) that 
$[Q,P]\,|\psi\>_{phys}=i\,|\psi\>_{phys}=0$, namely, that 
$|\psi\>_{phys}=0$ and thus ${H}_{phys}=\emptyset$, which
 is unacceptable. (This is a typical case of second-class 
constraints.) To avoid this situation, we replace the ideal
 conditions by the choice ${H}_{phys}\equiv\E{H}$, 
where in the present case $\E=\E(P^2+Q^2\le\hbar)=|0\>\<0|$,
 the projection operator onto the harmonic oscillator ground 
state, thus leading to a one-dimensional ${H}_{phys}$. 
More generally, we accommodate this kind of  situation by the 
criterion that ${H}_{phys}=\E{H}$, where
  \be \E=\E(X^2\le\delta(\hbar)^2)  \ee
and $\delta(\hbar)$ is an $\hbar$-dependent regularization 
parameter to be chosen on a case-by-case basis. The choice 
of a quadratic combination of constraints is not written
 in stone, and
a discussion of alternative choices is presented
 elsewhere \cite{k2}.

 The second manner in which the idealized situation may fail
 arises when the operator
$X^2\equiv\Sigma_{\alpha=1}^A\Phi^2_\alpha$ has its zero in 
the {\it continuous spectrum}. An example of this situation 
is given by the single classical constraint $\phi=q=0$, which
 then becomes the quantum constraint $\Phi=Q$. Other examples
 could arise from two degrees of freedom for which, in an
 obvious notation, e.g., (i) $\Phi_1=Q_1$, $\Phi_2=Q_2$, or
 (ii) $\Phi_1=Q_1$, $\Phi_2=P_2$, or (iii) $\Phi=Q_1^2+Q^2_2-1$,
 etc. In these cases $X^2\equiv\Sigma_{\alpha=1}^A\Phi^2_\alpha$
 has its zero in the continuum. As a consequence 
$\E(X^2=0)\equiv0$, which we deem to be unacceptable.
 In place of this idealized condition we once again choose
  \be \E=\E(X^2\le\delta(\hbar)^2)  \ee
where $\delta(\hbar)>0$. Observe, in the present case, that
 for any $\delta>0$, it follows that ${H}_{phys}=\E{H}$
 is an {\it infinite-dimensional} (regularized) physical
 Hilbert space. Although it is actually possible to work
 with this regularized space, say in cases where $\delta$ 
is extremely small, e.g., $\delta=10^{-1000}$, in some 
natural units, it is analytically preferable if we are
 able to take the limit $\delta\ra0$. However, this limit
 cannot be taken in any straightforward fashion since
   \be \lim_{\delta\ra0}\,
\<\lambda|\E(X^2\le\delta^2)|\phi\>\equiv0 \ee
for any pair $|\lambda\>$, $|\phi\>\in{H}$, whenever
 the zero of $X^2$ lies in the continuum.

The problem of quantum constraints with their zero in the
 continuum is well known and has often been studied in the
 literature. Briefly summarized, such problems have been
 studied, e.g., by (i) introducing Gel'fand triplets \cite{k3},
 (ii) specialized algebraic representations \cite{k4}, or (iii)
 rescaled limits within suitable subspaces \cite{k5}.

As one version of type (iii) above, we imagine working in a
 {\it representation} in which $X$ is diagonalized (as $x$),
 and therefore
  \be \<\lambda|\E(X^2\le\delta^2)|\phi\>=\int\int_{-\delta}^\delta
\,{\o{\lambda(x,y)}}\,\phi(x,y)\,dx\,d\sigma(y)\;, \ee
where the variable $y$ corresponds to any degeneracy. In this 
form it is clear, as $\delta\ra0$, that the right-hand side 
vanishes. To overcome this situation, let us first {\it restrict}
 attention to a subset of functions, e.g., 
  \be {\cal D}\equiv\{\chi(x,y):\,\chi(x,y)\in
\{{\rm polynomial}(x,y)\,e^{-(x^2+y^2)}\}\}\;,  \ee
and consider $\lambda=\lambda_0\in{\cal D}$ and 
$\phi=\phi_0\in{\cal D}$, so that \be \<\lambda_0|\E(X^2
\le\delta^2)|\phi_0\>\equiv\int\int_{-\delta}^\delta\,
{\o{\lambda_0(x,y)}}\,\phi_0(x,y)\,dx\,d\sigma(y)\;. \ee
In the present case it follows that 
\b &&\lim_{\delta\ra0}\,(2\delta)^{-1}\,\<\lambda_0|\E(X^2
\le\delta^2)|\phi_0\>\\
&&\hskip1cm=\lim_{\delta\ra0}\,(2\delta)^{-1}\int
\int_{-\delta}^\delta\,{\o{\lambda_0(x,y)}}\,\phi_0(x,y)\,
dx\,d\sigma(y)\\
&&\hskip1cm=\int\,{\o{\lambda_0(0,y)}}\,\phi_0(0,y)\,
d\sigma(y)\\
&&\hskip1cm\equiv (\lambda_0,\phi_0)\;.  \e
This final expression defines a sesqui-linear form 
characterizing a pre-Hilbert space. Completion of the
 pre-Hilbert space in the usual fashion [i.e., inclusion
 of limits of Cauchy sequences in the inner-product-induced
 norm $\|\phi_0\|\equiv\sqrt{(\phi_0,\phi_0)}$, plus the 
identification of elements as equivalence classes of 
functions as necessary] leads to the {\it true} physical 
Hilbert space, ${H}_{phys}$, in which the constraint
 condition $X^2=0$ is finally satisfied.

In the previous discussion it is noteworthy that we chose
a {\it representation} for the constraint operators in order
 to achieve a successful rescaling and limit as the
 regularization was removed. Other schemes may avoid 
the regularization and its subsequent removal, but they
 all require a {\it representation} to be introduced.

A natural question then arises whether or not it is possible
 to devise a procedure to enforce a quantum constraint whose
 zero lies in the continuum in an {\it abstract fashion}, i.e.,
 is it possible to impose such a constraint {\it without}
 introducing any specific representation whatsoever.
 It is this question that we address in the remainder of
 this paper, and it is noteworthy that the answer to that
 question is in the affirmative. 

The key to obtaining this affirmative answer is worth noting.
 The traditional view of obtaining a physical Hilbert subspace
 in the case of zero in the continuous spectrum is to {\it 
change the Hilbert space}, essentially, by imposing the 
constraint as a distribution. The new method introduced 
in this paper {\it changes the underlying topology} of 
the existing (pre-)Hilbert space and then is able to impose 
the constraint with a zero in the continuum in the same way
 one imposes a constraint with a discrete zero!

In this section, we have used the symbols $P$ and $Q$ in their
 traditional Heisenberg and Schr\"odinger senses. Note well
 that the same symbols $P$ and $Q$ appear in the subsequent
 sections, but they are used in a much wider sense. In
 particular, $Q$ is used as a {\it generic constraint operator
 with its zero in the continuum}, while $P$ is used as a {\it
 maximally symmetric operator} satisfying $[Q,P]=i\one$ on a 
suitable domain {\it which need not even be dense}.

\section{The continuous spectrum and topology}
Let us begin with the observation that, for topological 
reasons, in the case of zero being
in the continuous spectrum of $Q$, there cannot exist a projection from
the original Hilbert space $H$ to a physical Hilbert subspace $H_{phys}$: 
\sn
To see this, let us consider the simplest case, namely where the 
spectrum of the constraint operator $Q$ is nondegenerate, and is 
given by an interval, say $I=[0,\infty)$.
In the spectral representation of $Q$, the original
 Hilbert space $H$ is therefore 
the space $H=L^2(I)$ of square integrable functions
$f\in H$ over the interval $I$.
\sn
We would like to find a 
physical Hilbert subspace $H_{phys}$ corresponding
to ``$Q=0$". It should be a one-dimensional vector space 
of ``function values at $0$". We could then identify 
this space with ${\bf C}$.
To this end, let us consider the linear functional
$\phi$ defined by the property that
it maps every continuous function $f\in H$ onto its
value at zero, i.e.: 
\be
\phi: f \rightarrow f(0)
\ee
Intuitively, one might assume that
$\phi$ is the desired projection onto 
a one-dimensional ``physical"
subspace of function values at zero.  
\sn
 $ $ However, $\phi$ is not a projection.
The problem is that,
 even though 
we restricted the domain of $\phi$ 
to continuous representatives of  
square integrable functions,
$\phi$ is not a continuous map:
Consider, for example, the sequence 
$\{g_n(x)\}_{n\in \N}$ 
of continuous square integrable functions 
\be
g_n(x) := A~e^{-n x^2}.
\label{nullseq}
\ee 
for an arbitrary constant $A$. 
In the topology of the
Hilbert space $H$, namely the norm topology
 induced by the $L^2$ scalar product
of $H$, the sequence clearly converges to the null vector 
in $H$, i.e to the function $g(x)\equiv 0$ . Namely,
 $\lim_{n\rightarrow\infty}g_n=g$ in the sense that
$\lim_{n\rightarrow\infty}\vert\vert 
g-g_n\vert\vert^2 =\lim_{n\rightarrow\infty}
\int_{-\infty}^\infty dx~A^*A \exp(-2 n x^2)=0$.
On the other hand, $\lim_{n\rightarrow\infty}
\phi(g_n)=A$, which proves the discontinuity. 
\sn
This means the following: We can take any continuous square integrable
function $f\in H$ and add a null sequence such $g_n(x)$. Then, 
in the limit we recover
the vector $f\in H$. However, crucially, 
as far as the original topology of the Hilbert 
space $H$ is concerned, we can 
in this way give $\phi(f)$ any arbitrary value
that we wish, i.e. $\phi$ is ill-defined. 
\sn
Intuitively, the underlying reason for the ill-definedness of $\phi$
is of course that even continuous functions $f\in H$ can
be arbitrarily sharply peaked over a point, say zero, meaning that functions
 $f \in H$ could 
possess arbitrarily different values at zero -
 while their $L^2$-distance is arbitrarily small,
i.e. while such functions can still be arbitrarily close
within the topology of $H$.
\sn
Thus, the situation is that $\phi$ is
 noncontinuous and therefore also unbounded.
Since projectors are of course bounded, $\phi$ is 
not a projection! Another way to see this is that 
by Riesz' theorem, $\phi$, being noncontinuous,
is not contained in (the dual of) $H$, 
and can therefore not be projected onto.
\sn
The case of real physical interest is of
 course where the spectrum of $Q$ is 
degenerate at zero, and where one therefore 
expects the physical Hilbert space
to be higher dimensional. 
But it is clear that the fundamental problem, 
namely that $\phi$ is not a projector,
persists to all nontrivial cases - as long 
as we stick to the original topology of $H$. 
Our aim is therefore to suitably modify the topology of the original 
Hilbert space in order to be able to project onto the physical subspace.

\section{Hints from the theory of distributions and Sobolev spaces}

Before we describe our solution to the general problem, let us recall
some basic facts from the theory of distributions and
Sobolev spaces, facts which can  be directly
 used to solve the problem for the simple
nondegenerate  example
which we just discussed. Our general method 
will be motivated by this example.
\sn
We begin by recalling how, in spite of our arguments
above, the map $\phi$ can be made into a continuous and bounded map,
in fact, into a projection. 
The price to be paid is that the
{\it topology} of the original Hilbert space must be changed. 
Intuitively, the topology of the function 
space needs to be changed in such a manner that functions which
differ only by sharp peaks (which make
 arbitrarily little difference in the
$L^2$ topology) are separated in the new topology.
\sn
The basic observation underlying the theory of Sobolev spaces is that
a sufficiently improved topology is induced by a simple new scalar
product on the function space, see e.g. \cite{gwaiz}.
The new scalar product is arranged to be  
sensitive to the rate of change of functions. 
In the norm topology induced by this
scalar product even those functions can be separated which differ only 
by spikes that are so sharp that
the $L^2$ topology cannot distinguish them 
(e.g. those differing by sequences such as the 
sequence $\{g_n(x)\}$ of Eq.\ref{nullseq}).
\sn
In particular, while the scalar product of $H$ is of course
\be
\langle f_1\vert f_2\rangle = \int_I dx ~f^*_1(x) f_2(x)
\ee
the scalar product of the Sobolev Hilbert space $H^1$ is defined, using 
distributional derivatives, as:
\be
(f_1|f_2) = \int_I dx ~\left[ f^*_1(x) f_2(x)
+ \frac{df^*_1(x)}{dx} \frac{df_2(x)}{dx}\right]
\label{nscp}
\ee
Note that we will use throughout the notation 
$\langle\vert\rangle$ for the scalar
product in the initial Hilbert space and 
$(\vert)$ for the new scalar product.
\sn
Let us check whether the functional $\phi$ 
is a continuous
functional over $H^1$: To this end, we recall, by
Riesz' representation theorem, that continuous 
functionals over a Hilbert space
can be identified with the Hilbert space 
vectors themselves, via the
scalar product action. Thus, if $\phi$ is
 indeed a continuous functional
over $H^1$, then we should now be able to
 identify a representation of $\phi$ 
as a vector in $H^1$! In  fact, 
\be
\phi(x) = e^{-x}
\ee
is the representation of $\phi$ in $H^1$. It is normalized with respect to
the scalar product Eq.\ref{nscp} of $H^1$.
The reader may check in a short calculation that indeed  
$(\phi,f)=f(0)$ for all $f\in H^1$.
Thus, $\phi$, which maps continuous functions onto their values 
at a point, or more precisely
\be
\Pi~:=~ \vert \phi)~\otimes~(\phi \vert
\ee
becomes a projection, as desired. It is also clear that this could not have
been achieved without a change in topology.
\mn
{\bf Remark:  Higher Sobolev spaces}
\mn
\label{uni}
We should also mention that 
our particular choice of change of topology - to move from $H$ to $H^1$ -
is not the only possible one to achieve our goal of making $\phi$ a 
projection:
To see this, let us be more precise 
about the properties of the functions in $H^1$ as
opposed to those in $H$.
For a function $f(x)$ to be in $H^1$ it must not only be 
square integrable: $f(x)$ must also be the indefinite integral 
of some square integrable function ``$df(x)/dx$". We recall that
a function $f(x)$ is an indefinite integral exactly if it is absolutely 
continuous (on any countable set of non-overlapping intervals).
Thus, moving from the Hilbert space $H=L^2$
to the Hilbert space $H^1$ improves the continuity of the functions
 $f(x)$ to
absolute continuity. However,  
the derivatives $df(x)/dx$ of functions in $H^1$ need only be in $L^2$,
not $H^1$. 
Thus, in $H^1$ it is still not possible to project
$df(x)/dx$ onto its value at a fixed point. In fact, $df(x)/dx$ need only be
defined almost everywhere. As we are here only interested in projecting onto
the values of functions and not their derivatives this is not a problem for 
our purposes. 
\sn
Nevertheless, for completeness, let us recall how the topology could be
arranged to improve also the behavior of the functions' derivatives.
As the theory of Sobolev spaces shows, it is sufficient to this end to use 
higher derivative operators in the scalar product, i.e. 
to use e.g. the scalar product
$(f_1,f_2) := \int_I dx 
~\sum_{j=0}^n \frac{d^jf^*_1(x)}{dx^j} \frac{d^jf_2(x)}{dx^j}$
for some $n>1$ and its induced norm topology. In this way one obtains
the higher Sobolev spaces $H^n$. Obviously, $H\supset H^1 
\supset H^2 \supset H^3 ...$ $ $. 
In the function space $H^n$, we can project the 
derivatives $d^{m-1}f/dx^{m-1}$ of functions $f(x)$, up to $m=n-1$, onto 
their values at any given point. It is clear that,
 while we could use 
any higher Sobolev space, $H^1$ does suffice for our purposes here.

\section{The new method}

Let us now consider the general 
problem of projecting onto the physical 
subspace in the case of zero being in the 
continuous spectrum of a self-adjoint
constraint operator $Q$ whose spectrum may be arbitrarily degenerate.
We would like to generalize the procedure used when
zero is only in the point and not in the continuous
spectrum of the self-adjoint 
constraint operator $Q$, in which case 
one can straightforwardly 
define the physical subspace as the kernel of $Q$, i.e.
\be 
H_{phys}:= \mbox{ker}~Q=\mbox{ker}~Q^*\;.
\ee
In the case where zero is only in
the continuous and not in the point spectrum of $Q$, 
the problem is of course that 
then $\mbox{ker}Q=\mbox{ker}Q^*=\emptyset$, which is obviously not
the desired subspace. 
\sn
Our main idea in this paper is therefore to treat this case by
modifying the topology of the initial Hilbert space $H$
in the direction away from the ``constraint surface" in a manner analogous to
passing from an $L^2$ Hilbert space to the Sobolev space $H^1$. 
(In the general case, where zero is also in the point spectrum, one  
may project out the 
kernel of $Q$ as usual, before applying our procedure). 
\sn
The new Hilbert space, which we will call
$\tilde{H}$, will be a subspace of the original Hilbert
space when considered as a vector space, and 
the action of all operators, their commutation relations, etc, therefore 
remains unchanged. Also $Q$ is of course still the same, as a linear map.
Crucially, however, $\tilde{H}$ will be different as a Hilbert space,
being equipped with a new scalar product. This scalar product
changes the induced norm topology
in the direction of the constraint degree 
of freedom. As a consequence, the $*$ structure will change and the 
operator $Q$ will no longer be
self-adjoint, i.e. $Q\neq Q^*$! 
As our main finding, we will show that this modification
enables us to identify
the physical subspace as: 
\be
H_{phys} := \mbox{ker}~Q^*
\ee
Since $H_{phys}$ is therefore a proper Hilbert subspace,
 namely $H_{phys}\subset
\tilde{H}$,
this means that once one has passed to the new topology, 
i.e. from $H$ to $\tilde{H}$, one can 
again implement the constraint by projection.
\mn
Explicitly, in order to carry out the program of 
suitably modifying the topology
in the direction away from the constraint surface, 
we begin by completing the 
description of the degree of freedom which is to be constrained:
Namely, in addition to the constraint operator $Q$, let us also consider
a variable $P$ which is conjugate to $Q$, i.e. a maximal 
symmetric operator $P$
which obeys: 
\be
[Q,P]=i\one
\label{ccr}
\ee
By definition, therefore, the domain of $P$ is the
maximal domain $D_P\in H$ on which this commutation 
relation holds and
on which $P$ is symmetric. 
We remark that while in general $P$ 
will not be self-adjoint on $D_P$ - 
in particular it never is if $Q$ is positive -
the property of being maximally symmetric will suffice.
As we will see, it will also not matter for our purposes that 
the domain $D_{P}\subset H$ on which the commutation relation
Eq.\ref{ccr}  holds will in general not even be
dense in the original Hilbert space $H$ - as 
it clearly will not be if $Q$ also
possesses a point spectrum. Intuitively, this is
because for our purposes only
the part of the spectrum around zero matters. 
\sn
In analogy with the Sobolev space example for
functions, let us now consider the domain $D_{P^*}$ and let us change the
topology on it to obtain a new Hilbert space $\tilde{H}$:
Namely, as a vector space, we define 
$\tilde{H}$ to be identical to $D_{P^*}$ 
while we equip
$D_{P^*}$ with a new scalar product which then induces a
new norm topology.
Denoting the scalar product in the original Hilbert space $H$, and thus
on $D_{P^*}$, by $\langle~\vert~\rangle$, we define the scalar 
product $(~\vert~)$
on $\tilde{H}$ through:
\be
(v_1\vert v_2):=\langle v_1\vert v_2\rangle + 
\langle P^*v_1\vert P^* v_2\rangle
\label{newsp}
\ee
In other words, $\tilde{H}$ is the graph Hilbert space of $P^*$.
In the simple case where the spectrum of $Q$ is purely continuous and
nondegenerate
this Hilbert space is the Sobolev space $\tilde{H}=H^1$.
\sn
An advantage of our functional analytic definition is that, 
unlike in our discussion of the simple example of the nondegenerate spectrum
and its Sobolev space, we now no longer need to 
work in the spectral representation of $Q$. 
Let us denote the domain of $Q$ in $\tilde{H}$ by $\tilde{D}_Q$ (consisting
of all vectors $\phi\in \tilde{H}$
for which $Q\phi$ has finite norm with respect to the new scalar
 product given in Eq.\ref{newsp}).
On $\tilde{D}_Q$
the operator $Q$ is no longer
self-adjoint, $Q\neq Q^*$ (and not even symmetric).
We then define the physical Hilbert subspace
 $H_{phys}\subset \tilde{H}$ as 
the kernel of $Q^*$:
\be
H_{phys} := \mbox{ker} ~Q^* 
\label{defhphys}
\ee
This definition has a simple interpretation: 
We require that if we act with $Q$ on any
vector in its domain in $\tilde{H}$
then the resulting vector is orthogonal to all physical vectors.
In other words, we define the physical subspace $H_{phys}$ as the orthogonal
complement of the range of $Q$ in the Hilbert space $\tilde{H}$. 
\sn
For added clarity, let us be fully precise regarding
 the 
definition of $H_{phys}$ through Eq.\ref{defhphys}. As a vector space,
$H_{phys}$ is given by
\be
H_{phys} := \left\{ \vert v\rangle\in 
\tilde{H}=D_{P^*}
\vert ~ \forall w \in ~ \tilde{D}_Q~ : ~
\langle Q w\vert v\rangle + \langle P^* Q w\vert P^* v\rangle = 0\right\} 
\label{defhphys2}
\ee
i.e. all its vectors are also vectors in $H$,
 when considered as a vector space.
The scalar product in $H_{phys}$ is given by $(~\vert~)$ in Eq.\ref{newsp}.
\sn
To summarize, we begin by completing the picture
 of the degree of freedom which is to be
constrained, namely by augmenting $Q$ by a symmetric operator $P$ 
obeying the commutation relation $[Q,P]=i\one$ on its
 maximal domain $D_{P}$ in
the original Hilbert space $H$. Second, we change the scalar
 product and consequently the induced
topology on the domain $D_{P^*}\subset H$
to obtain the graph Hilbert space, $\tilde{H}$, of $P^*$
with the scalar product Eq.\ref{nscp}.
Third, we identify the physical subspace $H_{phys}$ as the
proper Hilbert subspace $H_{phys}= 
\mbox{ker}~Q^*$ of the Hilbert space $\tilde{H}$. 
Thus, after passing from $H$ to $\tilde{H}$ we can again project
onto a physical subspace which is a proper Hilbert subspace.
\mn
{\bf Remark:  The special case of positive $Q$}
\mn
If $Q$ is a
 positive self-adjoint operator with zero being in the
purely continuous spectrum, i.e. when zero is actually a 
boundary of a piece of the
continuous spectrum, then 
there is additional structure which can be used:
\sn
Namely, consider in this case $D_{P^*}$ modulo $D_P$
in the new topology, i.e.
consider $\tilde{H}_B:= D_{P^*}\ominus D_P$. By von Neumann's theory of
self-adjoint extensions of symmetric operators, the domains of $P^*$ and $P$
differ exactly by the space of boundary functionals,
and we have the von Neumann formula
\be
D_{P^*}\ominus D_P = N^+ \oplus N^-
\ee
where $N^\pm$ are the deficiency spaces of $P$.
 Since we assume that zero is a boundary of the
spectrum we can therefore conclude that $H_{phys}\subset 
N^+\oplus N^-$. As a consequence,
we can say that all physical vectors must be a linear 
combination of vectors in the kernels of
either the operator $(P^*+i)$ or $(P^*-i)$. In concrete 
representations, this fact can
yield useful differential equations.

\section{Examples}
Let us illustrate the working of the new 
method with simple examples in non relativistic
quantum mechanics. We will choose constraint operators
 $Q$ which are functions only of
the position operators $Q=Q(\x)$. In this way, the associated 
classical constraint surface, namely the zero set of
 $Q=0$ read as a classical
equation, will be obvious. We will show that our new
method projects on the Hilbert space of functions
 over the constraint manifold.

\subsection{Example: Projecting onto a point on the line}
As a first example, let us reconsider the simple case of a constraint $Q=0$ 
for a self-adjoint operator $Q$
whose spectrum is  
continuous at zero and nondegenerate. The nondegeneracy means of course
that we expect the physical Hilbert space to be one-dimensional.
This example will basically be the same as the one in which we discussed
Sobolev spaces. In order make this case slightly nontrivial 
let us choose the spectrum of $Q$ as: 
\be
I=[0,1]~\cup ~\{2,3,...\}
\ee 
Thus,  in the spectral representation of $Q$, our starting Hilbert space $H$
is $H= L^2(0,1)~\oplus~\bf l^2\rm, $ with scalar product:
\be
\langle \phi_1\vert \phi_2\rangle ~= ~
\int_0^1 dq ~ \phi_1^*(q) \phi_2(q) ~+~
\sum_{n=2}^\infty \phi_1^*(n) \phi_2(n)
\ee
We begin by considering a maximally symmetric
 operator which obeys $[Q,P]=i\one$,
namely $P=-i\partial_q$ on its domain $D_P$. Functions in $D_P$ are
absolutely continuous, square integrable, vanish at the interval boundary, 
and their almost everywhere defined derivative is also square integrable.
We note that, while $D_P$ is not dense
 in $H$ - since it lacks the eigenspaces to
the discrete eigenvalues - this will not matter
when we project onto ``$Q=0$". 
\sn
We now equip $D_P$ with the new scalar product, Eq.\ref{nscp},
to obtain $\tilde{H}$:
\begin{eqnarray}
(\phi_1\vert \phi_2) & := & \langle \phi_1\vert \phi_2\rangle + 
\langle P^*\phi_1\vert P^* 
\phi_2\rangle \\
  & = & \int_0^1 dq~ \left\{ \phi_1^*(q) \phi_2(q)~+~
\left(\partial_q  \phi_1(q)\right)^* \partial_q\phi_2(q)\right\} 
\label{15}
\end{eqnarray}
Next, we define the (Sobolev) Hilbert space $\tilde{H}$ as the graph
Hilbert space of $P^*$, i.e. as
 the domain of $P^*$, with the scalar product Eq.\ref{15}. 
\sn
Now we are ready to calculate the physical
 subspace $H_{phys}=\mbox{ker} ~Q^*$. 
Explicitly, according to Eq.\ref{defhphys2},
the condition for vectors $\psi\in \tilde{H}$ to be in the physical domain 
$H_{phys}$ now reads
\be
\int_0^1 dq~ \left\{ \phi^*(q) q \psi(q)~+~
\left(\partial_q q \phi^*(q)\right) \partial_q\psi(q)\right\} ~=
~ 0 \qquad  \forall ~\phi \in \tilde{D}_Q
\ee
This yields the condition, again for all $\phi \in D_Q$,
\be
\int_0^1 dq~ \left\{ \phi^*(q) q \psi(q)~-~
q \phi^*(q) \partial^2_q\psi(q)
\right\} ~+~\left[q \phi^*(q) \partial_q\psi(q)\right]_0^1 ~=~0
\ee
which means that
\be
q (\partial_q^2 -1)\psi(q) ~=~0
\label{c1}
\ee
and
\be
\partial_q\psi(q)\vert_{q=1}~=~0.
\label{c2}
\ee
There is only one solution to Eqs.\ref{c1},\ref{c2},
 up to normalization:
\be
\psi(q) ~=~ e^{-q} + e^{q-2}
\label{sol}
\ee
This vector spans the one-dimensional 
physical Hilbert subspace $H_{phys}\subset \tilde{H}$.
\sn
Let us check whether $\psi$ indeed projects 
functions onto their values at zero (up to
the normalization constant). Let $\phi$ be any 
vector in $\tilde{H}$. Indeed,
\begin{eqnarray}
(\psi\vert\phi) & = & \int_0^1 \left\{ \left(e^{-q} +e^{q-2}\right)\phi(q)
{}~+~ \left(\partial_q\left(e^{-q} +e^{q-2}\right)\right)
\partial_q\phi(q)\right\} \\
 & = & \left[\left(-e^{-q}+e^{q-2}\right)\phi(q)\right]_0^1\\
 & = & c ~\phi(0)
\end{eqnarray}
with the normalization constant $c=(1-e^{-2})$.
\sn
Finally, let us remark that since in this case
 $Q$ is positive, and zero is therefore 
a boundary of the continuous spectrum,
we could have found this 
solution also by using von Neumann's theory as mentioned above.
\sn
Namely, the only boundary vectors, i.e. the only vectors 
in $H$ obeying either 
$(P^*\pm i)\vert \psi\rangle =0$, i.e. the only normalizable solutions 
obeying  either of $(-i\partial_q \pm i)\psi(q)$, are $e^{\pm q}$.
 Thus, we could
have narrowed down the search for the physical subspace, 
knowing that it has
to lie within the boundary space $H_B$ spanned by these 
two vectors, as it does
of course, being spanned by $\psi$ in Eq.\ref{sol}.   

\subsection{Example: Similarly with isospinors}

Let us now consider an example of a constraint operator $Q$ whose spectrum
at zero is finitely degenerate, so that we can expect the 
corresponding physical,
i.e. constrained Hilbert space to be multi-dimensional.
\sn
To this end, let us consider the kinematics of a quantum mechanical particle
which possesses isospin and which lives, say, on the positive half 
line. In its
Hilbert space $H$ the scalar product of wave functions then reads:
\be
\langle\phi\vert\phi^\prime\rangle~=~
\sum_{i=1}^N\int_0^\infty dx~ \phi_i^*(x)\phi^\prime_i(x)
\ee
We wish to constrain the particle from the bulk to its boundary by choosing a
constraint operator $Q$, which acts as $Q\psi_i(x)=x\psi_i(x)$. 
By suitably
imposing ``$Q=0$" we intend to project wave functions onto
 the $N$-dimensional isospinor space at $x=0$ as the physical subspace. 
Within the
original topology of $H$ this is not possible because ker $Q = $ ker
 $Q^*=\emptyset$.
Following our general method, we therefore
 introduce the symmetric operator $P=-i\partial_x$,
obeying $[Q,P]=i\one$ on its domain. We can now equip this domain, or more 
precisely, the domain $D_{P^*}$, with the new scalar product Eq.\ref{newsp}
\be
(\phi\vert\phi^\prime)~=~\sum_{i=1}^N \int_0^\infty
 dx~\left\{\phi_i^*(x)\phi^\prime_i(x)
+(\partial_x\phi_i^*(x))\partial_x\phi^\prime_i(x)\right\}
\ee
to obtain the new Hilbert space $\tilde{H}$ which
 possesses the improved topology.
In $\tilde{H}$ the operator $Q$ is no longer self-adjoint and 
we identify $H_{phys}=\mbox{ker}~Q^*$.
According to Eq.\ref{defhphys2}, the 
condition for vectors $\psi\in \tilde{H}$ to be
in the physical subspace, i.e. in ker $Q^*$, now reads
\be
\sum_{i=1}^N\int_0^\infty dx~\left\{ \phi^*_i(x) 
x\psi_i(x) +(\partial_x x \phi_i^*(x))\partial_x \psi_i(x)\right\}~=~0
~~~\forall~\phi\in \tilde{D}_Q
\ee
which yields
\be
x(\partial_x^2-1)\psi_i(x)~=~0
\ee
The solution space is spanned by the wave functions 
$\psi_i^{(n)}(x)=\delta_{n,i} e^{-x}, (n=1,...,N),$ i.e. 
those vectors and their linear
combinations represent $H_{phys}$ in $\tilde{H}$. The 
projector $\Pi$ onto the physical
subspace is $\Pi = \sum_{n=1}^N \vert \psi^{(n)})\otimes 
(\psi^{(n)}\vert$.
\sn
Let us verify that the scalar product of the $\psi^{(n)}$ 
with an arbitrary wave function
$\phi_i(x)$ projects onto the isospinor space at zero.
Indeed,
\begin{eqnarray}
(\psi^{(n)}\vert\phi) & = & \sum_{i=1}^N\int_0^\infty dx~\left\{\delta_{n,i}
e^{-x} \phi_i(x) + \partial_x e^{-x}\delta_{n,i}\partial_x\phi_i(x)\right\}\\
 & = & \left[-e^{-x} \phi_n(x)\right]_0^{\infty} \\
& = & \phi_n(0)~.
\end{eqnarray}
Clearly, our treatment of this example straightforwardly also applies in the 
case $N=\infty$, to obtain an
infinite dimensional physical sub Hilbert space 
spanned by the $\{\phi^{(n)}\}$.
This means that the classical constraint ``manifold" which was here 
a set of $N$ discrete points could also be taken to be an infinite 
set of discrete points.

\subsection{Example: Projecting onto the boundary of the half-plane}
\label{half-plane}
Let us now consider
 the case where the classical constraint manifold is actually
a continuous manifold. To this end we consider the example of
 a particle which lives, say,
in a two-dimensional space on the half plane defined by $x_1\ge 0$, 
i.e. in its 
Hilbert space the scalar product of wave functions reads
\be
\langle \phi\vert\phi^\prime\rangle~=~ \int_{-\infty}^\infty dx_2~
\int_0^\infty dx_1~\phi^*(x_1,x_2)\phi^\prime(x_1,x_2).
\ee
We choose $Q=x_1$, in order to constrain the particle from
 the two-dimensional
bulk to its one-dimensional boundary at $x_1=0$. 
Clearly, $Q$ is positive, self-adjoint
and possesses the half-axis $x_1\ge 0$ as its infinitely degenerate spectrum.
\sn
To employ our method, we use the symmetric operator
$P := -i\partial_{x_1}$. As required, it obeys $[Q,P]=i\one$. 
Using $P$, the new scalar product reads
$(~\vert~ )$ 
\be
(\phi\vert \phi^\prime) ~=~ \int_{-\infty}^\infty dx_2~
\int_0^\infty dx_1~\left\{\phi^*\phi^\prime
+(\partial_{x_1}\phi^*)\partial_{x_1}\phi^\prime\right\}
\ee
which yields for our purposes
the topologically improved Hilbert space $\tilde{H}$ over
$D_{P^*}$. According to Eq.\ref{defhphys2}, physical 
vectors $\psi\in H_{phys}=$ ker $Q^*$
now obey
\be
\int_{-\infty}^\infty dx_2\int_0^\infty dx_1~\left\{\phi^* x_1\psi+
(\partial_{x_1}x_1\phi^*)\partial_{x_1}
\psi\right\}~=~0~~~~\forall~\phi\in \tilde{D}_Q,
\ee
which means:
\be
x_1 (\partial_{x_1}^2-1)\psi ~=~ 0
\ee
Thus, the physical vectors $\psi\in H_{phys}$ 
are represented in $\tilde{H}$ as wave functions
 $\psi(x_1,x_2)= f(x_2)e^{-x_1}$ where
$f(x_2)$ is an
 arbitrary square integrable function.
Let us explicitly verify that the scalar product of an arbitrary 
vector $\phi\in \tilde{H}$ with a
physical vector $\psi\in H_{phys}\subset \tilde{H}$ is the 
integral over the constraint
surface. Indeed:
\begin{eqnarray}
(\psi\vert\phi) & = & \int_{-\infty}^\infty dx_2 \int_0^\infty dx_1\left\{
f^*(x_2) e^{-x_1} \phi(x_1,x_2) + (\partial_{x_1} 
f^*(x_2) e^{-x_1})\partial_{x_1}\phi(x_1,x_2)
\right\}\nonumber \\
  & = & \int_{-\infty}^\infty dx_2 \left[-e^{-x_1}
 f^*(x_2)\phi(x_1,x_2)\right]^\infty_0\nonumber\\
 & = & \int_{-\infty}^\infty dx_2~ f^*(x_2) \phi(0,x_2)
\end{eqnarray}

\subsection{Example: Projecting onto a cylinder in $\R^3$}
For  a less trivial example, let us now consider quantum mechanics 
in three space-time dimensions, choosing as the constraint operator:
\be
Q = (X_1^2 +X_2^2-R^2)
\ee
We then of course expect the constrained, physical
 Hilbert space to be the Hilbert space
of square integrable functions over
the cylinder with radius $R$ around the $x^3$ axis. 
\mn
In the original Hilbert space, $H$, the scalar product 
of wave functions reads, choosing 
cylindrical coordinates,
\be
\langle \phi_1\vert \phi_2\rangle = 
\int_{-\infty}^\infty\int_0^{2\pi}\int_0^\infty dx_3~d\varphi
~dr~r~ \phi_1^*~\phi_2
\ee
while $Q$ acts as
\be
Q~\phi(r,\varphi,x_3) ~=~ (r^2-R^2)~\phi(r,\varphi,x_3)
\ee
As the first step, we introduce the operator
 $P:=-i \frac{1}{2r}\partial_r$ on its domain
$D_{P}$ of absolutely continuous square integrable functions $\phi$
 which vanish at $r=0$ and
for which the almost everywhere defined derivative
 $-i\frac{1}{2r}\partial_r\phi$
 is also square integrable.
Second, we equip $D_{P^*}$ with the new scalar product,
 Eq.\ref{nscp}, and its 
induced topology, to
obtain the Hilbert space $\tilde{H}$.
Third, we can now calculate the physical domain 
$H_{phys}=\mbox{ker} ~Q^*$.
According
to Eq.\ref{defhphys2}, a state $\psi$ is in
 the physical subspace exactly if for all $\phi
\in \tilde{D}_Q$ it obeys:
\be
0= 
\int_{-\infty}^\infty\int_0^{2\pi}\int_0^\infty dx_3~d\varphi
{}~dr~r~ \left(\phi^*(r^2-R^2) \psi
+ \left[\frac{1}{2r}\partial_r(r^2-R^2)\phi^*\right]
\left[\frac{1}{2 r}\partial_r\psi\right]\right)
\ee
\be
= 
\int_{-\infty}^\infty\int_0^{2\pi}\int_0^\infty dx_3~d\varphi
{}~dr~r~ \left(\phi^*(r^2-R^2) \psi
- (r^2-R^2)\phi^*\frac{1}{2r}\partial_r
\left[\frac{1}{2 r}\partial_r\psi\right]\right)
\ee
\be
+ 
\int_0^{2\pi}\int_{-\infty}^\infty d\varphi~dx_3\left[
\frac{1}{2} (r^2-R^2)\phi^*\frac{1}{2 r}\partial_r\psi\right]_0^{\infty}
\ee
Thus, physical states must obey the boundary condition 
at $r=0$, for all $\phi\in \tilde{D}_Q$,    
\be
\lim_{r\rightarrow0^+}~\int_0^{2\pi}\int_{-\infty}^\infty dx_3~d\varphi~
\frac{1}{2} (r^2-R^2)\phi^*\frac{1}{2 r}\partial_r\psi ~=~0
\label{bc}
\ee
since
 the boundary term at infinity vanishes 
due to the square integrability. Also, 
for all positive $r$ with $r\neq R$ the physical states
must obey the differential equation:
\be
0 ~=~  r~ (r^2-R^2) \frac{1}{2r}\partial_r\frac{1}{2 r}\partial_r\psi
\label{de}
\ee
The solutions to this differential equation are of the form: 
\be
\psi(r,\varphi,x_3)~=~f(\varphi,x_3)~\left(A e^{-r^2} + B e^{r^2}\right) 
\label{solf}
\ee
Let us begin by considering the part of the solution in the region $r<R$: 
We know that $\psi$ must be continuous everywhere, and this is
nontrivial at the origin, $r=0$: Namely, for $\psi$ to be continuous at the
origin one needs that either $\psi(r=0,\varphi,x_3)=0$, or 
that $\psi(r=0,\varphi,x_3)$
is independent of $\varphi$. Thus, we obtain two different possible behaviors 
of physical states in the region $r<R$: The first set of physical states 
vanishes at the origin, i.e.,
\be
\psi(r,\varphi,x_3)~=~f(\varphi,x_3)~\left(e^{-r^2} -e^{r^2}\right)~. 
\label{sol1}
\ee
Here, a priori, $f(\varphi,x_3)$ is some arbitrary 
square integrable function. 
However, solutions must also obey the boundary condition Eq.\ref{bc}.
Since all $\phi$ either vanish at $r=0$,
 or are $\varphi$-independent, Eq.\ref{bc}
takes the form:
\begin{eqnarray}
0 & = & \lim_{r\rightarrow 0^+}~\int_0^{2\pi}~d\varphi~
\frac{1}{2 r}\partial_r\psi\\
& = & \lim_{r \rightarrow 0^+}~\int_0^{2\pi} d\varphi~
f(\varphi,x_3)~
\left(-e^{-r^2} -e^{r^2}\right).
\end{eqnarray}
Thus,
\be
0 =   - 2  \int_0^{2\pi} d\varphi~f(\varphi,x_3)  \label{nozm}
\ee
which means that in solutions of the form given in Eq.\ref{sol1} the
$f(\varphi,x_3)$ are indeed arbitrary square integrable function - with
the exclusion of the $\varphi$-zero-modes.
\sn
However, the zero-modes are not lost: The solutions which we have just
obtained are only the set of solutions which are continuous at $r=0$ by
virtue of vanishing there. As we said above, there is a second set of
physical states, namely those which 
are continuous at $r=0$. These do not depend on 
the variable $\varphi$, i.e.
they are of the form
\be
\psi(r,\varphi,x_3)~=~g(x_3)~ \left(A e^{-r^2} + B e^{r^2}\right) ~.
\ee
For these solutions the boundary condition Eq.\ref{bc} reads
\begin{eqnarray}
0 & = & \lim_{r \rightarrow 0^+}~\int_0^{2\pi}~d\varphi~
\frac{1}{2 r}\partial_r\psi\\
& = & \lim_{r \rightarrow 0^+} \int_0^{2\pi} d\varphi~
g(x_3)~ \left(-A e^{-r^2} +B e^{r^2}\right)\\
& = & -2 \pi g(x_3) (A-B) 
\end{eqnarray}
which implies that $B=A$. Thus, the physical states
which carry the zero-modes of the angular degree of freedom $\varphi$
are represented in the region $r<R$ by the functions:
\be
\psi(r,\varphi,x_3) ~=~ g(x_3)~ \left(e^{-r^2} +e^{r^2}\right)
\ee
Clearly, for the part of the solution in the region $r>R$, 
square integrability 
requires solutions of the form of Eq.\ref{solf} with $B=0$.
 The solutions of the differential
equation for $r<R$ and for $r>R$
are glued together by the requirement that every physical state 
is continuous at
$r=R$. We therefore obtain that the physical states are 
spanned by states of one of two forms. Either,
\be
\psi(r,\varphi,x_3) ~=~ f(\varphi,x_3)~
\left\{\begin{array}{ll} (e^{r^2} - e^{-r^2}),~~~~ r<R\\ \\
e^{-r^2}~(e^{2 R^2}-1), ~~~~r\ge R
\end{array} \right.
\ee
where $f$ is an arbitrary square integrable 
function of $x_3$ and $\varphi$, without,
however, the zero-modes
 in $\varphi$, because of Eq.\ref{nozm}.
The rest of the physical states are the zero-modes in 
$\varphi$, and are represented
as functions of the form:
\be
\psi(r,\varphi,x_3) ~=~ g(x_3)~\left\{\begin{array}{ll}
 (e^{-r^2} + e^{r^2}),~~~~ r<R\\ \\
e^{-r^2}~(1+e^{2 R^2}), ~~~~r\ge R
\end{array} \right.
\ee
The physical Hilbert space $H_{phys}\subset 
\tilde{H}$ is spanned by these functions, 
and is 
equipped with the scalar product of $\tilde{H}$, as given in Eq.\ref{newsp}.
\sn
In order to project an arbitrary function in $\tilde{H}$ down to $H_{phys}$
we need to take scalar products of arbitrary functions
$\phi$ in $\tilde{H}$ with functions
in the physical Hilbert subspace $H_{phys}$. Let us check that, as desired,
this scalar product reduces to an integral over the product of two 
functions over the surface of the cylinder $r=R$:
\begin{eqnarray}
(\phi,\psi) & = &
\int_{-\infty}^\infty\int_0^{2\pi}\int_0^\infty dx_3~d\varphi
 {}~dr~r~ \left(\phi^*\psi
+ \left[\frac{1}{2r}\partial_r\phi^*\right]
\left[\frac{1}{2 r}\partial_r\psi\right]\right)
\\ 
& = & 
\int_{-\infty}^\infty\int_0^{2\pi}\left\{\int_0^R dx_3~d\varphi
{}~dr~r~ \left(\phi^*\psi
- \phi^*\frac{1}{2r}\partial_r\left[\frac{1}{2 r}\partial_r\psi\right]\right)
 \right. \label{diff1}
\\
&  & ~~~~~~~~~+~\left. \int_R^\infty dx_3~d\varphi 
{}~dr~r~ \left(\phi^*\psi
- \phi^*\frac{1}{2r}\partial_r\left[\frac{1}{2 r}\partial_r\psi\right]\right) \label{diff2}
\right\}\\
&  &  +   
\int_0^{2\pi}\int_{-\infty}^\infty d\varphi ~dx_3~\left\{\left[
\frac{1}{2} \phi^*\frac{1}{2 r}\partial_r\psi\right]_{0_+}^{R_-} + \left[
\frac{1}{2} \phi^*\frac{1}{2 r}
\partial_r\psi\right]_{R_+}^\infty \right\}\label{bc3} 
\end{eqnarray}
where, to be precise, in line \ref{bc3} the 
evaluations at the interval boundaries 
are limits taken from within the interval. 
The terms in lines \ref{diff1} and \ref{diff2} vanish
 because all physical states $\psi$
obey Eq.\ref{de}. In order to evaluate line \ref{bc3} we decompose
$\phi$ into its $\varphi$- zero-modes $\phi_0$ and the rest, $\phi_1$:
\be
\phi(r,\varphi,x_3)~ = ~ \phi_0(r,x_3) + \phi_1(r,\varphi,x_3)
\ee
where $\phi_1(r=0,\varphi,x_3)=0$. Recall that we found that while 
the physical subspace
consists of all square integrable functions, say 
$h(\varphi,x_3)$ over $x_3$ and over $\varphi$-space, its 
representation as functions in the original Hilbert space realizes the
$\varphi$-zero-modes somewhat specially. Namely, if we decompose 
an arbitrary 
$h(\varphi,x_3)$ into
their $\varphi$-zero-modes and the rest as 
\be
h(\varphi,x_3)=f(\varphi,x_3)+g(x_3),
\ee
then these are represented as
\be
\psi(r,\varphi,x_3) ~=~ \left\{\begin{array}{ll} 
f(\varphi,x_3)~(e^{r^2} - e^{-r^2}) + 
g(x_3)(e^{r^2}+e^{-r^2}),~~~ r<R\\ \\
f(\varphi,x_3)~e^{-r^2}~(e^{2 R^2}-1)+ 
g(x_3) e^{-r^2}(1+e^{2R^2}), ~~~r\ge R
\end{array} \right.
\ee
We then read off: 
\begin{eqnarray}
\frac{1}{2r} \partial_r\psi\vert_{0_+} & = & 2 f(\varphi,x_3) \\
\frac{1}{2r} \partial_r\psi\vert_{R_-} & = & 
 f(\varphi,x_3)(e^{R^2}+e^{-R^2})~+~g(x_3)(e^{R^2}-e^{-R^2}) \\
\frac{1}{2r} \partial_r\psi\vert_{R_+} & = & 
f(\varphi,x_3)(-e^{R^2}+e^{-R^2})~-~g(x_3)(e^{R^2}+e^{-R^2}) 
\end{eqnarray}
Using
\be
\phi_1(0,\varphi,x_3)~=~0,
\ee
\be
\int_0^{2\pi}d\varphi~ \phi_1^*(r,\varphi,x_3)g(x_3) ~=~0
\ee
and
\be
\int_0^{2\pi}d\varphi~ \phi_0^*(r,x_3)f(\varphi,x_3) ~=~0
\ee
we therefore find that:
\begin{eqnarray}
(\phi,\psi) & = &
\frac{1}{2}\int_0^{2\pi}\int_{-\infty}^\infty d\varphi ~dx_3~\left\{
\left(\phi_0^*(r,x_3)+\phi_1^*(r,\varphi,x_3)\right)\frac{1}{2r}
\partial_r\psi\vert_{r=R_-}\right.\\
 &  & ~~~~~~~~~~~~~~~~-\left(\phi_0^*(r,x_3)+
\phi_1^*(r,\varphi,x_3)\right)\frac{1}{2r}\partial_r\psi\vert_{r=0_+}\\
 &  & ~~~~~~~~~~~~~~~~-\left. 
\left(\phi_0^*(r,x_3)+\phi_1^*(r,\varphi,x_3)\right)\frac{1}{2r}
\partial_r\psi\vert_{r=R_+}\right\}\\
 & = & e^{R^2}~ 
\int_0^{2\pi}\int_{-\infty}^\infty d\varphi ~dx_3~
\left\{\phi_0^*(R,x_3)g(x_3)~+~
\phi_1^*(R,\varphi,x_3)f(\varphi,x_3)\right\}\\
 & = & e^{R^2} 
\int_0^{2\pi}\int_{-\infty}^\infty d\varphi ~dx_3~
\phi^*(R,\varphi,x_3) h(\varphi,x_3)
\end{eqnarray}
as it should be, with the factor $e^{R^2}$ being an overall 
normalization constant.

\subsection{Example: Projecting onto a line in $\R^3$}
Let us briefly also consider the case of projecting not onto a cylinder, but
onto a one-dimensional line in $\R^3$. 
To this end, we consider, similar to the previous
example:
\be
Q~=~x_1^2+x_2^2 
\ee
and again $P=-i\frac{1}{2r}\partial_r$, where $r^2=x_1^2+x_2^2$.
Within our formalism, the change of the dimensionality 
of the constraint manifold
is automatically taken care of: The
 differential equation obeyed by the physical 
states, Eq.\ref{de}, must now hold for all
 positive $r>0$. Thus, normalizability
shows that all solutions must
be of the form: $\psi(r,\varphi,x_3)=f(\varphi,x_3)e^{-r^2}$.
 Since these functions
do not vanish at the origin, $r=0$, continuity at $r=0$
 requires independence of $\varphi$:
\be
\psi(r,\varphi,x_3) ~=~ g(x_3) e^{-r^2}
\label{spsol}
\ee
While for $R>0$ these solutions were ruled out by the 
boundary condition Eq.\ref{bc},
here Eq.\ref{bc} is trivially obeyed due to its prefactor $r$.
\sn
Let us check whether the scalar product of any physical state $\psi$
with an arbitrary state $\phi$ reduces to the integral of the product
of the two functions over the $x_3$ axis. Indeed,
\begin{eqnarray}
(\phi\vert \psi) & = & \int_{-\infty}^\infty dx_3\int_0^{2\pi}d\varphi~
\left[\frac{1}{2}\phi^*(r,\varphi,x_3)\frac{1}{2r}\partial_r
\psi\right]_0^\infty \\
& = & \int_{-\infty}^\infty dx_3\int_0^{2\pi}d\varphi~
\frac{1}{2}\phi^*(0,\varphi,x_3)g(x_3)\\
 & = & \pi \int_{-\infty}^{\infty} dx_3 ~\phi^*(0,x_3) g(x_3)
\end{eqnarray}
where in the last step we used that the
functions $\phi$ do not depend on $\varphi$ at $r=0$.
Thus, the scalar product of arbitrary functions in $\tilde{H}$
with a function from the physical subspace indeed 
yields  the integral of the product of the
two functions over the $x_3$ axis, as expected.
\sn
This example also shows how the formalism takes care of 
a dimensional reduction: 
The loss of the $\varphi$-degree of freedom  
is expressed by the representatives of physical states
possessing only zero-modes of the $\varphi$-degree of freedom.
In other words, restricting the function space over a 
cylinder to only its zero-modes in
the angle $\varphi$ is the same as reducing the function space to
a function space over a line rather than the cylinder.

\subsection{Example: An impossible constraint}
It is perhaps instructive to briefly discuss how 
an ``impossible" constraint yields
an empty physical subspace. To this end, let us 
consider again in $\R^3$ the constraint
\be
Q~=~x_1^2+x_2^2 +R^2 
\ee
for some positive $R>0$.
In this case, physical states must obey the differential 
equation Eq.\ref{de} for all
positive $r>0$ and therefore, by the above arguments,
 the only possibility are
functions of the form given in Eq.\ref{spsol}. However, 
unlike in the case $R=0$,
now the boundary condition Eq.\ref{bc} is nontrivial, namely yielding
$g(x_3)=0$. This rules out any solutions, thereby yielding
$H_{phys}=\emptyset$, as should be the case. 

\section{Uniqueness of $Q$ and $P$, and the
 topology of the constraint manifold}
As we saw already in Sec.\ref{uni}, our method cannot be unique:
While the topology of the Sobolev space $H^1$ is good enough 
to allow the projection onto a function's value at a point,
higher Sobolev spaces $H^n$ could be used as well. 
Analogously, in our more general method, for a given choice of $Q$ and $P$,
the topology of $\tilde{H}$ allows us to project onto the physical
Hilbert subspace, while higher Sobolev space analogs of $\tilde{H}$ 
could be used as well.
Interestingly, however, there is even more non-uniqueness in our
general method.
\sn
Firstly, 
 for a given classical constraint manifold
there is a non-uniqueness in the possible choices of the 
constraint operator $Q$.
Secondly, also
for a given $Q$ there  exists in general
 a non-uniqueness in the possible
choices of a canonical conjugate operator $P$.
The question in how far the various physical Hilbert subspaces
obtained from these different choices are or are not equivalent
is subtle and reveals interesting relations to the 
dimensionality and topology
of the constraint manifold. 
\subsection{Nonuniqueness of $Q$ for a given classical constraint}
Perhaps surprisingly, even simple classically identical constraints 
may differ quantum mechanically, even when there are 
no ordering ambiguities:
In order to see this, 
let us consider the example of the constraint operator 
 $Q_1=x_1^2+x_2^2+x_3^2-R^2$ in
the Hilbert space $H$ of square integrable functions in $\R^3$.
Classically, the constraint manifold, i.e. the
set of solutions to $Q_1=0$ is a 2-sphere in $\R^3$: The 
physical Hilbert space which we obtain from $Q_1$ is of course 
the $L^2$ Hilbert space of functions over this sphere. 
Similarly, the constraint operator
$Q_2=x_3^2 + (x_1^2+x_2^2-R^2)^2$ describes the equator of the 
sphere described by $Q_1$.
\sn
Now, let us consider the constraint operator $Q_3 := Q_1 Q_2$. 
Classically, its constraint manifold, i.e. the set of points for which
$Q_3=0$, is still the 2-sphere with radius $R$, i.e. it is
 identical to the case above.
Quantum mechanically, the physical Hilbert space for $Q_2$,
however, can now be different from the space of functions over the sphere.
This is because $Q_2$ counts the 1-sphere of the equator twice. 
We might expect that the physical Hilbert space also specifically contains
$L^2$ functions over the equator and their scalar product. In this way,
the scalar product could be
consisting of evaluating both the 
two-dimensional integral over the sphere and the
one-dimensional integral over the equator of the
 product of two wave functions. 
\sn
This issue will be further investigated elsewhere.
It is clear that whenever such ambiguities arise, i.e. whenever the
physical subspaces can be, for example, 
function spaces over a variety of topologically 
different manifolds (of varying,
and in general possibly not even well-defined dimensionality), then
this should also have a reflection in the functional analysis of $P$.
 
\subsection{Non-uniqueness of the choice of $P$ for a given $Q$}
For a given choice of 
$Q$, our condition that $P$ be a maximal symmetric operator
 obeying $[Q,P]=i\one$
does not determine the choice of operator $P$ uniquely.
Clearly, any two operators $P$ and $P^\prime$
which obey the commutation relation $[Q,P]=i\one$
will differ by a symmetric operator $R$ which commutes with $Q$,
i.e. $P^\prime=P + R$ with $[R,Q]=0$, with the appropriate domains
understood. 
Usually, two such operators $P$ will be connected by a gauge
transformation, in the sense that
there exists a unitary operator $U$ in $H$, such 
that $P^\prime=P+R=UPU^\dagger$.
In this case, as is easily verified, $U$ is also an
 isometry connecting the corresponding 
``generalized Sobolev spaces" $\tilde{H}$ and $\tilde{H}{}^\prime$, 
as well as being an isometry connecting the
corresponding 
 physical subspaces $H_{phys}$ and $H_{phys}^\prime$. 
In this case, the choice of $P$ is therefore immaterial. 
\sn
For example, if two operators $P$
differ by an operator $R$ which is a self-adjoint operator that is
independent of the constraint degree of freedom, in the sense that it
commutes with $Q$ and $P$, then
we find $U=\exp(i Q R)$. Or, if $R$ is a function $k(Q)$ of $Q$
 with integral
$\int dq ~ k(q) =K(q)$, then we have $U=\exp(i K(Q))$.
\sn
More concretely, let us reconsider the example of Sec.\ref{half-plane}, where
we used $Q=x_1$ to project onto the boundary 
(the $x_2$-axis) of the half-plane $(x_1\ge 0)$.
We had chosen $P=-i\partial_{x_1}$. But this choice was not
unique because obviously e.g.
also $P^\prime := -i\partial_{x_1} + g(x_2)$ obeys $[Q,P^\prime]=i\one$.
The question arises as to whether we obtain the same physical Hilbert space
if we use $P^\prime$ rather than $P$. A short calculation shows that, using
$P^\prime$, the physical vectors are represented as functions
$f(x_2) e^{-x_1(1+ig(x_2))}$. The difference therefore is only a phase,
which vanishes for $x_1=0$, i.e. after the projection. Thus, the use of
$P^\prime$ yields the same Hilbert space as does the use of $P$. 
\sn
Similarly, also e.g. $P^{\prime\prime}:= -i\partial_{x_1}-i\partial_{x_2}$
obeys the commutation relation $[Q,P]=i\one$.
Using $P^{\prime\prime}$, physical vectors must obey the
differential equation $(\partial_{x_1} +\partial_{x_2} -1)\psi=0$.
Using the coordinates $z:=(x_1+x_2)/2$ and $y:=(x_1-x_2)/2$, we obtain
$(\partial_z-1)\psi=0$, yielding for the general expression of physical
vectors: $\psi= f(-y)e^{-z}$, i.e. $\psi=f(x_2-x_1)e^{-(x_1+x_2)}$.
The scalar product of a general vector $\phi \in \tilde{H}$ with a physical
vector $\psi\in H_{phys}$ is then:
\begin{eqnarray}
(\psi\vert\phi) & = & \int_0^\infty dx_1 \int_{-\infty}^\infty dx_2
\left\{ \psi^* \phi + \left( (\partial_{x_1}+\partial_{x_2})f^*(x_1-x_2)
e^{-(x_1+x_2)/2}\right)(\partial_{x_1}+\partial_{x_2})\phi\right\}\nonumber\\
 & = & 
\int_{-\infty}^\infty dx_2 \left[ f^*(x_2-x_1) e^{-(x_1+x_2)/2}\phi
\right]_0^{\infty}\nonumber\\
 & = & \int_{-\infty}^\infty dx_2 f^*(x_2)e^{-x_2/2}\phi(0,x_2)\nonumber\\
 & = & \int_{-\infty}^\infty dx_2 \psi^*(0,x_2) \phi(0,x_2)
\end{eqnarray} 
Thus, when using $P^{\prime\prime}$ 
we still obtain the same physical Hilbert subspace.
\sn
Similarly, also $P^{\prime\prime\prime}:= -i\partial_{x_1} +g(x_1)$ 
obeys the commutation relation. Its use leads to physical vectors
being represented as functions $\psi= f(x_2)e^{-x_1 - i\int^{x_1}g(x) dx}$.
All we obtain is a new phase factor
which vanishes (up to possibly an irrelevant global overall phase) after the
projection onto $x_1=0$. Thus, we obtain the same Hilbert space
also from $P^{\prime\prime\prime}$, as long as $g(x_1)$ is integrable. 
\sn
Even though these examples illustrate the robustness of the projection onto
the physical subspace under changes of $P$ we cannot exclude that,
in general, $R$ may not be a pure gauge, i.e. it may not be
``integrable" in these simple ways. 
One may find unitarily non-equivalent
maximal representations of the
commutation relation $[Q,P]=i\one$. On the other hand, 
while a priori these representations 
will lead to non-equivalent generalized Sobolev 
spaces, some of 
the resulting representations on the corresponding physical subspaces 
may still be equivalent after projection. 
\mn
The investigation of the precise 
relation between different choices of $Q$ for the same classical
constraint, unitarily non-equivalent $P$'s, and the non-uniqueness
of the topology of the ``constraint manifold" is probably deep and 
should be worth pursuing further.
\mn
\bf Acknowledgement: \rm The authors are happy to thank B. G. Bodmann
and S. Shabanov for useful comments.


\end{document}